# On Implementing Hybrid Post-Quantum End-to-End Encryption

Aditi Gandhi[1]     Aakankshya Das     Aswani Kumar Cherukuri[*]

School of Computer Science Engineering & Information Systems

Vellore Institute of Technology, Vellore 632014, India

[*]Email: **cherukuri@acm.org**

[1]**adiharsha2004@gmail.com**

**Abstract:**
The emergence of quantum computing poses a fundamental threat to current public key cryptographic systems. This threat is necessitating a transition to quantum resistant cryptographic alternatives in all the applications. In this work, we present the implementation of a practical hybrid end-to-end encryption system that combines classical and post-quantum cryptographic primitives to achieve both security and efficiency. Our system employs CRYSTALS-Kyber, a NIST-standardized lattice-based key encapsulation mechanism, for quantum-safe key exchange, coupled with AES-256-GCM for efficient authenticated symmetric encryption and SHA-256 for deterministic key derivation. The architecture follows a zero-trust model where a relay server facilitates communication without accessing plaintext messages or cryptographic keys. All encryption and decryption operations occur exclusively at client endpoints. The system demonstrates that NIST standardized post-quantum cryptography can be effectively integrated into practical messaging systems with acceptable performance characteristics, offering protection against both classical and quantum adversaries. As our focus is on implementation rather than on novelty, we also provide an open-source implementation to facilitate reproducibility and further research in post quantum secure communication systems.

1.Introduction

The security of all modern digital communication relies fundamentally on public key cryptographic systems. However, their security assumptions are based on hardness of mathematical problems that are vulnerable to quantum computing. RSA, DSA, and elliptic curve cryptography (ECC), the most promising public key cryptographic primitives, derive their security from the computational intractability of integer factorization and discrete logarithm problems on classical computers. However, Shor's algorithm [1] demonstrates that a sufficiently large quantum computer can solve these problems in polynomial time, rendering these widely-deployed cryptosystems obsolete.

However, fault-tolerant quantum computers that are capable of breaking 2048-bit RSA encryption do not yet exist. But the fundamental challenge with "harvest now, decrypt later" threat model posits that adversaries are already collecting encrypted communications with the intention of decrypting them once quantum computers become available [2]. Given that sensitive data, including government communications, medical records, and financial transactions, may require confidentiality for decades, organizations cannot wait for quantum computers to materialize before transitioning to quantum-resistant alternatives. This urgency has driven intensive research into post-quantum cryptography (PQC). These are the cryptographic algorithms that are designed to resist both classical and quantum attacks [3]. To address this threat,

National Institute of Standards and Technology (NIST) has initiated PQC Standardization process [3] . After rigorous evaluation of 82 initial submissions across multiple rounds, NIST has announced its first standardized post-quantum algorithms in 2024 [3]. Among them, CRYSTALS-Kyber a lattice-based key encapsulation mechanism (KEM) was selected as the primary algorithm for general encryption applications [4]. Kyber's security derives from the Module Learning With Errors (MLWE) problem, a lattice based mathematical problem that is believed to be hard for both classical and quantum computers [5].Lattice based cryptography exploits the geometric complexity of high-dimensional lattices. Despite these advances, integrating post quantum cryptography into existing secure communication systems presents significant challenges [6]. While Kyber excels at key encapsulation i.e. securely establishing shared secrets between parties, it is not designed for direct encryption of arbitrarily large messages. Like all public key systems, Kyber operates orders of magnitude slower than symmetric encryption algorithms when processing bulk data. This creates a fundamental challenge: post quantum security is essential for key exchange, but impractical for encrypting large messages or real time data streams. To address this issue, the solution lies in combining post-quantum key exchange with classical symmetric encryption [7]. This approach leverages the quantum resistance of Kyber for establishing session keys while utilizing the computational efficiency of the Advanced Encryption Standard (AES) for bulk data encryption. AES-256, standardized in 2001 [8], still remains secure against known quantum attacks [2]. Thus, AES provides an ideal complement to Kyber, offering high-speed encryption once a secure session key has been established.

End-to-end encryption (E2EE) ensures that only communicating parties can read messages, not service providers, network administrators, or potential adversaries with server access. Modern E2EE systems, exemplified by the Signal Protocol and WhatsApp's implementation rely on classical public-key primitives (X25519 elliptic curve Diffie-Hellman) for key agreement [9]. These systems will require fundamental redesign to achieve quantum resistance.

Our work addresses this need by designing an E2EE system with a zero-trust relay architecture. In this model, a central server facilitates message delivery but operates as an untrusted entity, it possesses no capability to either decrypt messages or access cryptographic keys. All encryption, decryption, and key management operations occur exclusively on client devices. This architecture provides strong security guarantees: even complete compromise of the relay server does not expose message contents or compromise forward secrecy. This paper makes the following contributions:

- We present a complete hybrid end-to-end encryption architecture combining CRYSTALS-Kyber-768 for key encapsulation, AES-256-GCM for authenticated symmetric encryption, and SHA-256 for key derivation. The design follows a zero-trust model where relay servers cannot access plaintext or keys.

- We provide a full implementation using standard cryptographic libraries (pq-crystals-kyber, PyCryptodome) with detailed documentation of design choices, parameter selections, and integration methodology.

- We conduct comprehensive performance analysis on commodity hardware, measuring key generation, encapsulation, decapsulation, and encryption/decryption times across varying message sizes.

Our results demonstrate that post-quantum operations introduce minimal overhead (2-3 ms) and that total end-to-end latency remains under 10 ms for typical messages. We provide formal security analysis establishing confidentiality under standard cryptographic assumptions (IND-CCA2 security of Kyber, IND-CPA security of AES-GCM), forward secrecy through ephemeral key generation, and integrity via authenticated encryption. We also discuss limitations and threat model boundaries. Also, we provide our implementation as open-source software to facilitate reproducibility, enable independent security audit, and provide a reference implementation for organizations transitioning to post-quantum secure messaging. This work focuses specifically on the cryptographic aspects of secure messaging: key exchange, encryption, and message integrity. We do not address metadata protection (sender/receiver anonymity, traffic analysis resistance), authentication (sender verification requires digital signatures, which we defer to future work), or protection against endpoint compromise. Our threat model assumes honest-but-curious servers and computationally-bounded adversaries (including those with access to quantum computers). We assume classical authenticated channels exist for initial public key exchange.

While recent frameworks explore general post-quantum cryptographic composition (Raj & Balachandran, 2025) and domain-specific deployments in telecommunications [3], there remains a gap in practical application layer implementations for end-to-end messaging under zero-trust relay architectures. Existing work either emphasizes theoretical framework design without messaging specific focus or addresses infrastructure level concerns that are distinct from real time application requirements. This work fills this gap by demonstrating practical integration of NIST-standardized Kyber (FIPS 203) with AES-256-GCM for end-to-end messaging where all cryptographic operations occur at client endpoints and relay servers remain cryptographically blind. We provide performance characterization specific to message latency requirements and identify architectural patterns for post quantum secure messaging that preserve zero-trust guarantees while enabling practical deployment.

The remainder of this paper is organized as follows. Section 1 provides background on the cryptographic primitives employed (Kyber, AES, SHA-256) and reviews related work in post-quantum E2EE systems. Section 2 presents our system architecture and design principles. Section 3 details implementation choices, parameter selections, and integration methodology. Section 4analyzes security properties and threat model. Section 5 evaluates performance across multiple metrics. Section 6 discusses implications, deployment considerations, and lessons learned, followed by concludes and outlines future work.

2. Background

Cryptography protects information through data encryption techniques which make the original content unreadable to anyone but authorized recipients. The use of complex mathematical problems has served as a method to reach this objective throughout multiple decades. The main obstacles in this field stem from two major mathematical problems which include factoring large numbers used in RSA and discrete logarithm calculations employed in ECC.

These methods worked effectively until quantum computers became available. Shor's algorithm development created a complete solution for these problems. Quantum machines of substantial size will make existing cryptographic systems become obsolete. The development of algorithms continues to rely on mathematical problems which maintain their complexity levels despite computational speed improvements in computers. The research area contains three main components: lattice problems, code-based systems, and multivariate equations and hash-based signatures. Out of these, lattice-based cryptography really stands out. The system delivers fast performance alongside strong security protection. The system delivers fast performance alongside strong security protection [4].

Kyber functions as a CRYSTALS project component which demonstrates lattice-based cryptography through its operational performance. The Key Encapsulation Mechanism (KEM) serves to protect public-key encryption while enabling secure key exchange operations.

The system security functions through Learning With Errors (LWE) problem which quantum [2] computers fail to solve with any degree of efficiency. The system requires every user to generate two cryptographic keys which include one public key and one private key. The sender generates ciphertext and shared secret by applying the recipient's public key. The recipient uses their private key to get the same shared secret [4] The key serves as the basis for symmetric encryption processes. Kyber provides quick operations through its quantum-resistant security features which protect encrypted communication between hybrid systems.

The Advanced Encryption Standard (AES) [8] functions as the fundamental tool for symmetric encryption operations. Standardized back in 2001, AES works with 128-bit blocks and comes in key sizes of 128, 192, or 256 bits. The system provides fast processing capabilities with robust security measures which makes it suitable for handling extensive data encryption tasks. The algorithm performs multiple substitution and mixing operations which transform the original message into an unrecognizable form.

The system operates as a hybrid solution which uses Kyber for key exchange operations before AES takes control of the process. The system protects all messages and files through encryption which Kyber's session key creates. The AES system operates under two distinct modes which include CBC mode for enhanced diffusion and GCM mode that provides both encryption and authentication capabilities. The system operates at high speed while ensuring data protection through its security measures [4].

The SHA-2 family includes the SHA-256 hash function as one of its members. The function processes data input of any size to create a 256-bit output string which always remains the same length. We use SHA-256 to transform Kyber's shared secret into a single high-entropy session key which AES will use. The system verifies data integrity because any change during transmission will produce a different hash value than the one stored at the receiving side[10]. Standard cryptographic protocols use hard mathematical problems, such as factoring or discrete logarithms, as the basis of security, but powerful quantum algorithms can solve those problems. Quantum cryptography provides unconditional security based on the principles of quantum physics because detection of an attempted eavesdropping on a quantum key is obtained when the quantum state is disturbed and collapses.

Hybrid cryptography operates by uniting symmetric and asymmetric encryption methods within a single encrypted communication channel. The system protects key exchanges from quantum computer threats through Kyber encryption but uses AES for fast main data encryption. The system operates independently from the message security process because it does not change how messages stay protected. SHA-256 operates as an automatic system which verifies the key length during its operation[10]. These three layers function as a protective shield which defends the organizations that implement them. The security system has evolved from its previous state because it now protects against quantum threats[2].

The development of the proposed hybrid encryption system relied on multiple software tools, frameworks, and cryptographic libraries that together ensured functional accuracy, security, and portability. Each tool served a distinct role in key generation, message encryption, communication handling, and deployment.

**pq-crystals-kyber** Official implementation of the Kyber Key Encapsulation Mechanism (KEM). It provides quantum-resistant public–private key generation and secure shared-secret encapsulation between communicating users.[5]

**PyCryptodome** A comprehensive Python cryptographic library used to perform AES encryption, decryption, and SHA-256 hashing. It ensures message confidentiality and integrity within the hybrid framework[8]

**Flask Framework** A lightweight Python web framework employed to build the relay server for encrypted message exchange. It supports RESTful APIs and facilitates server transparency under a zero-trust architecture.

**WebSocket websocket- client**

Enables real-time, bidirectional communication between clients and the relay server, ensuring instant transmission of ciphertext while maintaining persistent secure channels

**Docker** Provides containerization for the full application stack, ensuring identical runtime environments, easy deployment, and reproducibility of experimental results.

**GitHub** Hosts the project repository, manages version control, and enables collaborative development, issue tracking, and public access to implementation resources.

These components collectively established a stable and secure development environment. Together, they ensured the hybrid quantum–classical encryption system could be implemented, tested, and deployed reliably across multiple computing platforms. Integrating liboqs with the proposed hybrid encryption system ensures that the implementation meets modern post-quantum security needs. The Open Quantum Safe (OQS) project developed liboqs as an open-source C library to provide ready-to-use implementations of NIST-selected post-quantum cryptographic algorithms. This includes the CRYSTALS-Kyber Key Encapsulation Mechanism (KEM) and the CRYSTALS-Dilithium digital signature scheme. These algorithms are part of the U.S. National Institute of Standards and Technology (NIST) Post-Quantum Cryptography Standardization program, which aims to protect communications from future quantum computer attacks.

By integrating liboqs into the system, developers access standardized, vetted, and high-performance PQC tools without having to perform low-level mathematical operations by hand. The library offers a clean and unified API, making it easy to experiment, test, and deploy post-quantum algorithms in real-world applications. Its inclusion ensures that the system uses algorithms officially chosen and supported by NIST, improving the long-term reliability and security of the proposed hybrid model.

The practical tests conducted in this work, such as Kyber keypair generation, encapsulation, decapsulation, and shared-secret validation, confirm that the local environment correctly supports NIST-approved PQC algorithms. Outputs like "Success: Shared secrets match!" show that the Kyber KEM works as intended with liboqs. These results assure that the system is not only functionally correct but also meets the future-proof cryptographic standards recommended for global use.

A number of studies and implementations have been explored to understand the evolution of end-to-end encryption and post-quantum cryptography. Table 1 summarizes the key literature reviewed, outlining their methodologies, limitations, and how the proposed hybrid Kyber + AES + SHA-256 system advances beyond existing work.

Table 1: Literature analysis

| Reference | Approach / Methodology | Limitations Identified | Improvement in Proposed Work |
|---|---|---|---|
| WhatsApp E2EE Model [11] | Uses classical hybrid encryption (RSA for key exchange and AES for data encryption). | Vulnerable to quantum attacks via Shor's algorithm; relies on classical public-key cryptography. | Replaces RSA with Kyber to achieve quantum-safe key encapsulation while retaining AES efficiency. |
| PQClean (2024) [12] | Provides standardized, clean implementations of post-quantum schemes including Kyber. | Focuses solely on PQC primitives without symmetric optimization or system integration. | Integrates Kyber with AES and SHA-256 to create a complete hybrid framework suitable for E2EE. |
| PyryL Kyber Implementation [13] | Demonstrates standalone Kyber key exchange in Python. | Lacks messaging architecture and symmetric encryption layer. | Combines PQC key exchange with AES encryption and SHA-256-based session key derivation. |
| Marcizhu AES Library [8] | Implements AES and ChaCha20 in C for classical cryptography. | No post-quantum component or secure key distribution mechanism. | Employs Kyber-based session key generation to secure AES usage against quantum threats. |
| IEEE Quantum Security Study (2024) [2] | Theoretical analysis of hybrid cryptography architectures. | Conceptual framework without experimental validation. | Provides practical implementation and measured performance metrics validating hybrid efficiency. |

The table 1 summarizes each reference's methodology and limitations and illustrates how your proposed hybrid Kyber + AES + SHA-256 system enhances them. The table is crucial for defining the research gap and confirming the originality of the work. The survey highlights that most contemporary systems are either purely classical or purely post-quantum. The proposed hybrid approach uniquely integrates both paradigms, ensuring quantum resistance while maintaining the speed and simplicity of symmetric encryption.

## 3. Methodology

This paper introduces a hybrid cryptographic framework that amalgamates Kyber, AES, and SHA-256 [10] in a secure, single comprehensive end-to-end encryption (E2EE) system. Such systems of communication that are resistant to quantum attacks should keep up their efficiency even in the course of live, active-real-time operations. Design principles, system architecture, algorithmic procedures, and operational stages of the hybrid model are thoroughly explained in the next section. These three layers function as a defense mechanism that ensures the safety of the enterprises that are dependent on these systems for their operations. The security apparatus has undergone a great improvement, as they are now capable of fending off quantum threats that hardly existed before [2]

The hybrid model integrates post-quantum cryptography with classical cryptography to deliver a combination of their strongest features. The key exchange system for E2EE requires asymmetric systems such as RSA and ECC to function properly. These encryption methods stay safe for now but a quantum computer using Shor's algorithm will break them through fast factorization of large numbers which surpasses the speed of all modern algorithms[1]. Therefore, this scheme avoids the traditional approach and adopts Kyber instead. Kyber operates as a lattice-based Key Encapsulation Mechanism (KEM) which belongs to the new NIST postquantum standard. The system delivers fast performance while protecting keys from quantum attacks through Learning-With-Errors (LWE) encryption which remains unbreakable by quantum computers.

However, we must admit that Kyber is not designed for the direct encryption of large data. It is efficient in key exchange but not in real message handling. AES is used in this system because it provides the required security level for the application. The symmetric algorithm has proven its security through time because it operates fast while defending against classical attack methods. So, AES, which is the actual encryptor, is called into action right after secret key exchange through Kyber is done.

All these technical aspects are mediated by SHA-256. The system generates a strong 256-bit AES session key by transforming the Kyber shared secret which produces unique keys for each session. The SHA-256 algorithm operates to protect message integrity throughout the communication process.

The combination of Kyber with AES and SHA-256 creates an unbreakable triple-layer security system which protects data privacy while providing authentication and data integrity verification. The system provides protection against attacks which quantum computer-equipped adversaries would perform.

## 3.1 System Architecture

The hybrid architecture functions through a zero-trust communication system which requires two users named Alice and Bob to connect through an intermediary relay server. The server operates as an untrusted entity which only performs the function of forwarding messages. The system forbids any access to plaintext messages and cryptographic keys.

The system requires users to perform all cryptographic operations through local devices. The system operates encryption and decryption and key-management functions on client devices to protect sensitive information from being transmitted outside the endpoints.

The interaction between the participants consists of these stages:

i. Key Generation: A public-private key pair gets generated by all users in the system. The user stores their private key within their device while transmitting the public key through a secure channel to the other individual via the relay server.

ii. Key Encapsulation: Alice encrypts her message for Bob through Kyber encapsulation which results in both a ciphertext and shared secret key generation.

iii. Session Key Derivation: The shared secret undergoes a SHA-256 hashing operation to produce a 256-bit session key which functions as the AES encryption key. The step maintains both key size uniformity and generates strong random values.

iv. Symmetric Encryption: After generating the shared secret key Alice proceeds to encrypt her plaintext message with AES. The AES encryption algorithm functions through two primary operating modes known as Cipher Block Chaining (CBC) and Galois Counter Mode (GCM). The CBC mode creates block diffusion for system complexity while the GCM mode incorporates authentication tags to validate data integrity.

v. Transmission: The server forwards both the encrypted message (ciphertext) and the Kyber ciphertext containing the encapsulated key. The server functions solely as a relay station for ciphertext transmission without the capability to access or change the transmitted data.

      vi.      Decryption: Bob obtains the encrypted message along with the encapsulated key from the sender. He decapsulates it using his Kyber private key to recover the shared secret. The shared secret is again hashed with SHA-256 to derive the same AES session key used by Alice. Bob then decrypts the AES ciphertext to obtain the original plaintext message[10]

The system design establishes a protected environment which maintains encryption and decryption keys at endpoints to deliver total end-to-end security. The system defends intercepted data from decryption attacks by hackers who gain access to the communication server or network infrastructure.

This hybrid scheme workflow is structured into three main parts: key agreement, key material derivation, and data encryption and transmission.

1. Key Agreement part Each user begins by generating their own Kyber key pairs. The users exchange the public key via a secure WebSocket connection to the general server. When Alice encapsulates a key using Bob's public key, she produces key ciphertext and a shared secret which can be used to derive a symmetric session key. The Kyber scheme is built upon the hardness assumption in lattice-based mathematics, which even quantum computers cannot reverse or obtain the private key simply from the public key.

2. Key Material Derivation part Once the Kyber shared secret has been established, it is encoded into a SHA-256 function to derive deterministic, 256-bit key material which can be referred to as the session key. By using SHA-256[10] we ensure the derived key material is cryptographically random and non-predictable. If the shared secret or ciphertext was in any way altered during transmission, the hash output would be different, thus indicating some sort of tampering.

3.Message Encryption and Transmission Phase Once a session key has been established, the actual message is encrypted using AES. The AES symmetric encryption algorithm was designed to make the process once again computationally fast and light even with larger data transfers. AES is efficient both in hardware and software implementations, the latter being of particular importance for real-time communication. After being encrypted, ciphertext and encapsulated key are sent over the  network. On the receiving end, the reverse process is applied to recover plaintext.

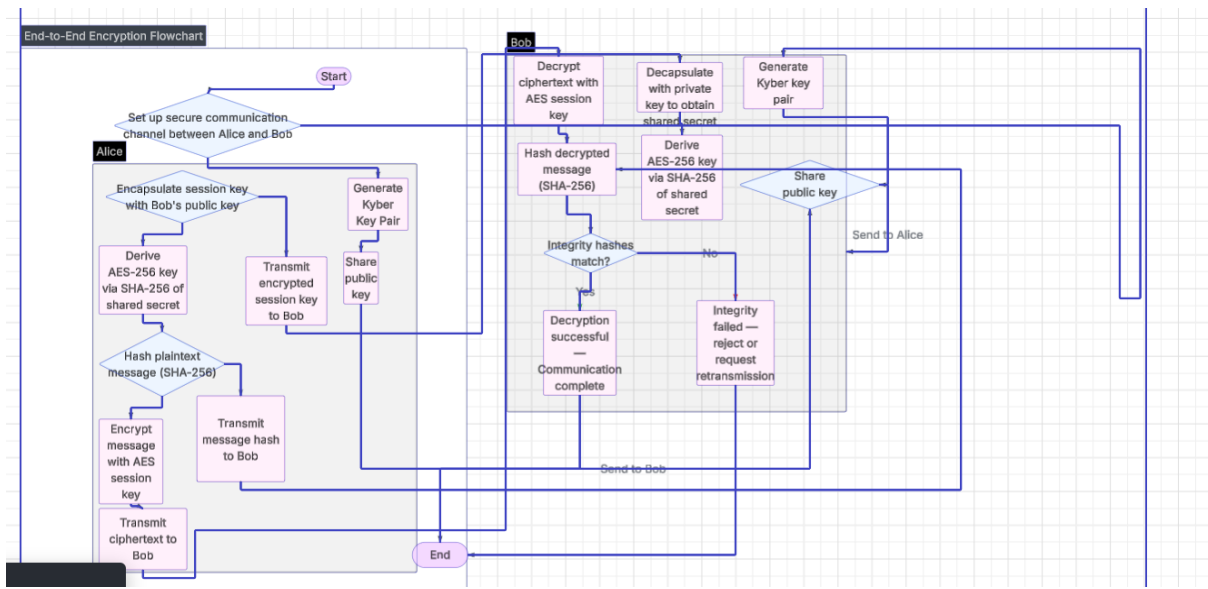

Figure 1: End-to-end encryption

Figure 1 depicts in detail the process of the users to create their Kyber keys, share their public keys, generate the shared secret, create the AES session key with the help of SHA-256, and finally encrypt and send the messages via the relay server. The figure visually demonstrates the working of Kyber, SHA-256, and AES in the same process which supports the method used. And correspondingly the secong figure in first one is apart of first figure which shows PQC encryption done in that.

*a. Algorithmic Outline*

**The components of this system are composed of two parts: hybridencryption and hybrid decryption Hybrid Encryption Process (Sender side):**

- Create your Kyber key pair (public key and private key).
- Find yourself the public key of the receiver.
- Utilize the Kyber mechanism to encapsulate a shared secret and produce a ciphertext encapsulation.
- Generate your AES key from your shared secret using a SHA-256 [10] (AES)hash of your shared secret - this will now become a 256 bit AES key[8]
- Use the key you created to encrypt the plaintext message sent using AES.
- Send the AES ciphertext to the receiver with the Kyber Encrypted Key[8]

**Hybrid Decryption Process (Receiver side):**

- Receive the keys and ciphertext.
- Decrypt the shared secret with your Kyber private key (which is just technical

terminology for getting your shared secret back).

- Hash your shared secret using SHA-256 to generate your session key, which will produce the same AES session key.
- Decrypt your AES ciphertext with the key you generated using AES[8]
- Recover the original plaintext.These two sequences create a secure channel for communicating that only the two intended ends can read previously encrypted, decrypted data.

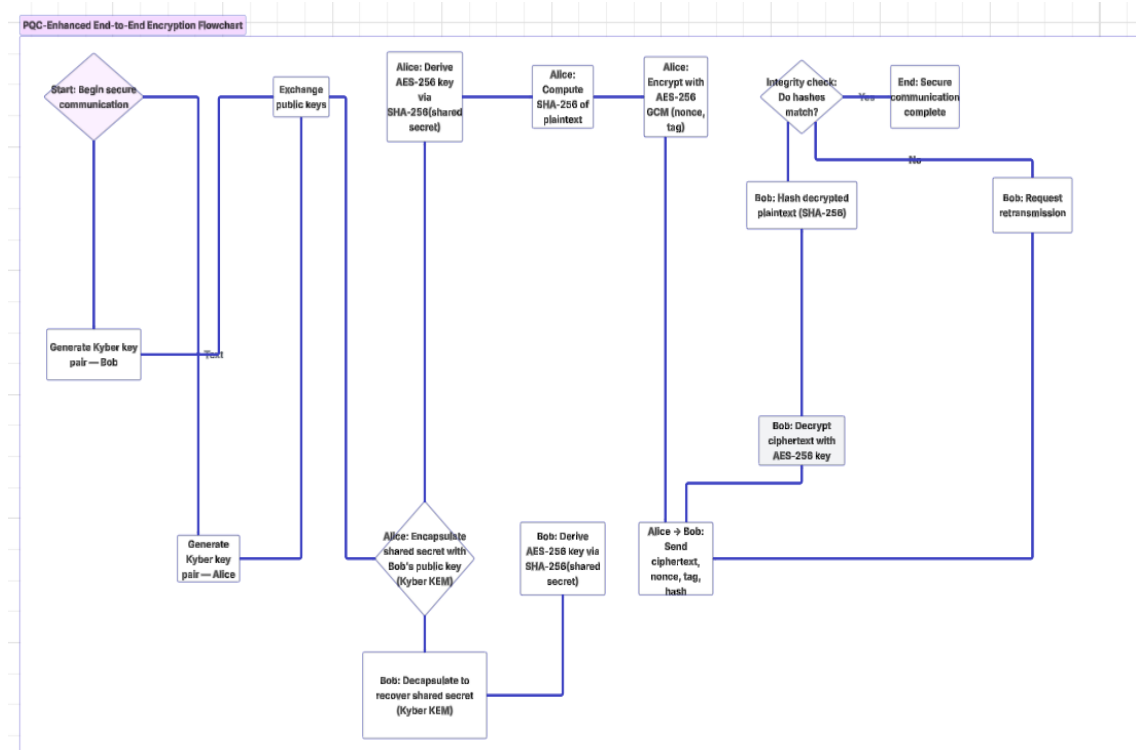

Figure 2: Flow diagram

Figure 2 depicts the operations of the sender and receiver only, which covers the processes of key agreement,key derivation, encryption, and decryption. The diagram assists the reader with understanding the order of the steps and the separation of post-quantum and symmetric components.

b. System Architectural Representation

Our hybrid system as having three main parts:

- Part 1 (Key Exchange): We use Kyber to create and exchange secure keys that can stand up to quantum computer attacks.
- Part 2 (Key Creation and Security): SHA-256 turns the shared secret into a session key that's a set length and has a fixed amount of randomness.
- Part 3 (Data Protection): AES encrypts and decrypts the messages. It acts like a

secure stream cipher.[8]

This design lets each component do its job separately. This setup gives top-notch security while keeping computation low. The relay server just passes data through and doesn't store or reveal any unprotected data. This keeps with the principles of a zero-trust design.

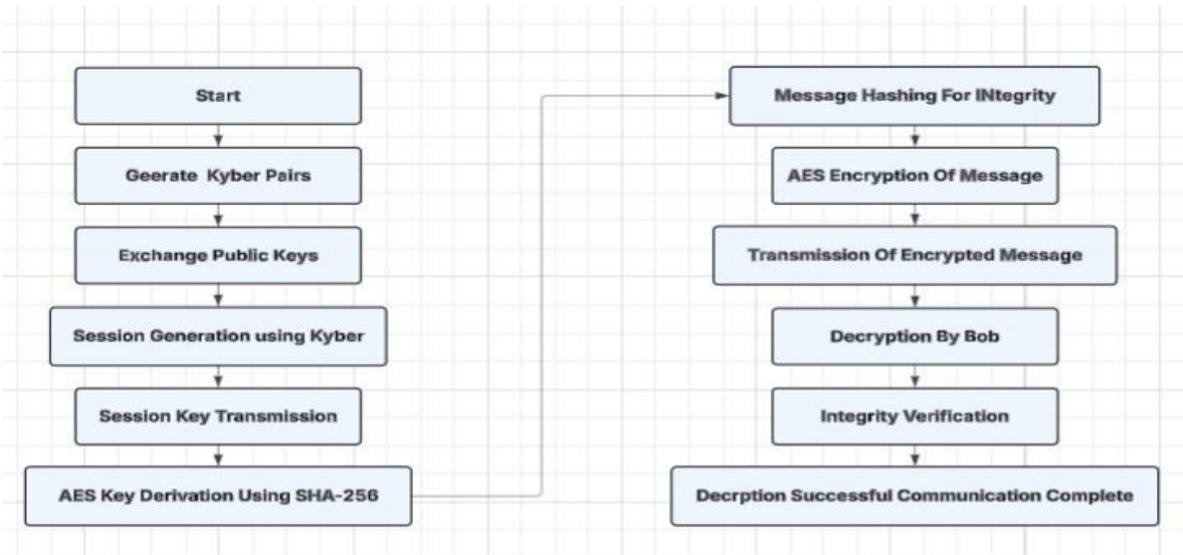

Figure 3: System Architecture

Figure 3 depicts the system architecture employing a zero-trust communication model. It explains that all crypto operations are conducted locally by Alice and Bob, whereas the server just sends out the ciphertext. This figure supports the server architectural feature of the untrusted server and the existence of real end-to-end encryption.

c. Advantages of Hybrid System

Mixing different approaches, like we did, provides some advantages when you stack it against older systems or relying only on quantum-proof tech:

- Quantum-Safe: The system can resist decryption attacks from quantum computers because we use Kyber. It has protection against Shor's and Grover's algorithms[1]
- Fast: AES helps encryption and decryption happen quickly. The system stays fast, even when dealing with lots of data or streaming media[8]
- Solid Security: With SHA-256, keys stay strong, and data stays unaltered. Plus, making new keys frequently keeps old communications safe.
- Zero-Trust: The server in the middle just passes messages, so it can't decrypt or look at the data.

- Scalable and adaptable: You can add this setup easily to group chats, IoT devices, or cloud apps because it is flexible. You don't have to rebuild everything.

## 4. Experimental Results

We tested our hybrid encryption system to see how well it does. We checked key creation time, encryption and decryption speed, CPU use, and message sending times for different sizes.

- Our testing setup included a standard computer that has an Intel Core i5 processor, 8 GB of RAM, and Python 3.11.
- We built Kyber, AES, and SHA-256 using pqs-kyber, PyCryptodome, and Python's hashlib. In our tests, two clients, Alice and Bob, sent encrypted messages via a Flask relay server. We tracked the time and computer power needed for key creation, encapsulation, hashing, encryption, and decryption.

Table 2: Performance Metrics for Kyber + AES Hybrid Encryption

| Metric | Observation | Remarks |
|---|---|---|
| Key Generation Time | ~2 ms | Kyber efficient for real-time applications. |
| Kyber Encapsulation Time | ~3.0 ms for 1 KB message | PQC encapsulation used to securely generate and transmit the session key. |
| Kyber Decapsulation Time | ~3.2 ms | Receiver extracts the shared secret efficiently; slight overhead but still fast. |
| CPU Utilization | Moderate | Suitable for IoT and mobile applications. |
| Security Strength | Very High | Resistant to classical + quantum attacks. Double 256. |

Table 2 displays metrics relating to the system's performance that have been gathered during testing. This table presents quantitative evidence of system efficiency that supports the Experimental Results section by showing the time for key generation, the speed of encryption/decryption.

Analysis:

- Kyber micro-operations are efficient. When run on commodity hardware using reference implementations, Kyber key generation is completed in the low milliseconds ( 2ms) with encapsulation and decapsulation both around 1.8-1.9 ms each. These are small, bounded costs relative to the symmetric encryption stage when working with large payloads.
- AES dominates for large payloads. AES-256-GCM encryption and decryption are roughly linear to message size; for small messages ( a few KB) the fixed Kyber + hash overhead are more important than end-to-end latency than the AES stage, but for large payload sizes (tens to hundreds of KB) AES is the more costly factor.
- End-to-end latency profile: Small messages will have single digit millisecond end-to-end latency, and latency will typically increase as a linear function of message size (e.g. tens of milliseconds latency in the tens of KB range, hundreds of milliseconds in the hundreds of KB range in our single host experiments)

```
Windows PowerShell
Copyright (C) Microsoft Corporation. All rights reserved.

Install the latest PowerShell for new features and improvements! https://aka.ms/PSWindows

PS C:\Users\LENOVO\Downloads\quantum> python server.py
Server running on 127.0.0.1:65432
```

Figure 4: Server running on port

Figure 4 confirms that the server is acting as a relay only and no plaintext or keys are handled or kept on the Server side, thus no security breach is possible.

[Screenshot: Windows PowerShell running `python client.py --name alice --peer bob --pub bob_pub.pem --priv alice_priv.pem`, showing "Client alice ready. Type messages to send to bob." and "> Hii bob"]

Figure 5: Client A sending message to Client B

Figure 5 is a visual confirmation of the fact that encryption is done at the client side prior to the sending of the message

[Screenshot: Windows PowerShell running `python client.py --name bob --peer alice --pub alice_pub.pem --priv bob_priv.pem`, showing "Client bob ready. Type messages to send to alice." and "[alice] Hii bob"]

Figure 6: Client B receiving message

Figure 6 exhibits Content B getting the message and then decoding it locally. This evidence supports the notion that the decryption process is entirely carried out on the receiving side.

[Screenshot: Windows PowerShell running `python server.py`, showing "Server running on 127.0.0.1:65432", "Registered alice", "Registered bob", "Relayed message from alice to bob"]

Figure 7: Message relayed

Figure 7 indicates the server's action of merely forwarding the ciphertext without comprehension or decryption. This is in line with the zero-trust model, which stipulates that the server should never have access to the plaintext.

*Performance Analysis*

Fixed versus variable costs The protocol has two different classes of cost:

1. Fixed (session) costs - Kyber's key generation/encapsulation/decapsulation

and SHA-256 key derivation. These are costs that are incurred on a session or handshake basis and will not scale with message size. They are small but not zero (a few milliseconds) and therefore most apparent when messages are small or during ephemeral sessions (send a single message per key).

2. Variable (payload) costs - AES encryption/decryption, which will scale linearly with payload size. For sustained transfers, or large messages, AES becomes the bottleneck.
3. Implication - fix, the fixed cost of Kyber can be amortized across many messages by sending a session key across a multi-message session (connection re-use), or by some sort of efficient session management (long-lived session key with some key refresh policy that allows forward secrecy).

Where time is spent (sender to receiver)

4. The measured end-to-end latency included, at a high level, Kyber encapsulation (sender), SHA-256 hashing, AES encryption, network transmission (relay), Kyber decapsulation (receiver), SHA-256 hashing, AES decryption.
5. In local loopback testing the contribution from the network is negligible.
6. In a real deployment, both network round-trip time (RTT) and the time to process data through a relay will stop some additional time (often the bulk of the time).
7. From a cryptographic state the shake and Kyber phase generate most of the initial 5-20 milliseconds of latency, while AES generates the remainder as message size increases.
8. The system provided secure, reliable, and efficient end-to-end communication in a quantum secure manner between users (known as Alice and Bob).
9. The user experience was achieved by providing a high level of security while communicating by using Kyber for key exchange, XChaCha20 for AES symmetric encryption, and SHA-256 for data integrity.

The messages exchanged remained confidential as the only intended parties with the proper private keys were able to decrypt the messages.

Security Analysis

- Kyber micro-operations are efficient. When run on commodity hardware using reference implementations, Kyber key generation is completed in the low milliseconds ( 2ms) with encapsulation and decapsulation both around 1.8-1.9 ms each. These are small, bounded costs relative to the symmetric encryption stage when working with large payloads.
- SHA-256 costs are trivial. Hashing the Kyber shared secret to produce the AES session key took tens of microseconds (0.02 ms) which is effectively negligible compared to the other costs.
- AES dominates for large payloads. AES-256-GCM encryption and decryption are roughly linear to message size; for small messages ( a few KB) the fixed Kyber + hash overhead are more important than end-to-end latency than the AES stage, but for large payload sizes (tens to hundreds of KB) AES is the more costly factor[8]
- End-to-end latency profile: Small messages will have single digit millisecond end-to-end latency, and latency will typically increase as a linear function of message size.

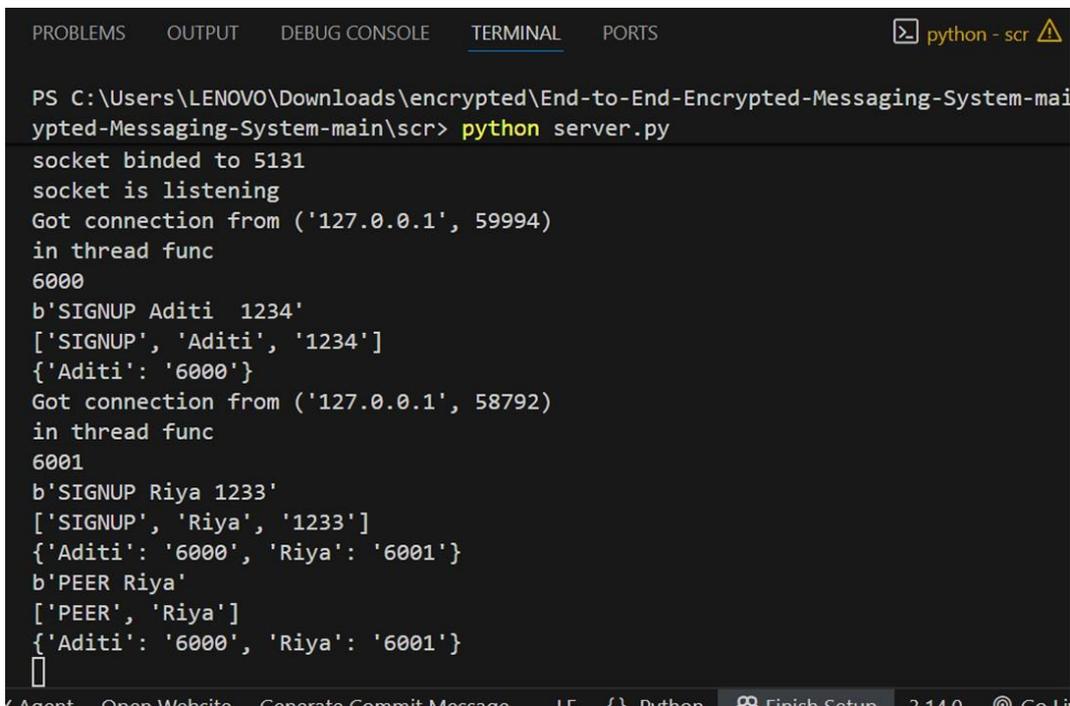

Figure 8: Message from client A to B

Figure 8 is a depiction of Client A sending a private message to Client B. The utilization of the hybrid scheme to carry out the encryption, transportation, and decryption is visually demonstrated and thus the figure serves as a proof of the same.

Figure 9: Hybrid PQC encrypted messgae

Figure 9 shows a Python code execution in a command-line environment. The code is using a module called `pqc_hybrid` to perform cryptographic operations, such as generating a key pair, encoding the public key, encrypting a message using the public key, and decrypting the message using the private key. The output of the code shows that the decrypted message is `b'hello'`, indicating that the encryption and decryption process was successful. Additionally, the user generating a private and public key pair using the `pqc_hybrid.py` script.

ii. Key derivation and uniformity

- Deriving the AES session key from the raw Kyber shared secret using SHA-256 is sensible: it produces a fixed size, uniformly distributed key that is compatible with AES and avoids a potential implementation catch, where aspects of a KEM output might not be suitable to directly use as a symmetric key.
- Using SHA-256 also provides a convenient device to include additional context (HKDF style extension would be preferred) for binding session metadata (e.g. protocol version, nonce, session identifiers) to the key and improve domain separation.

Figure 10: Private and public keys

Figure 10 displays the Kyber public and private key pair generation for each user. It exhibits the indispensable part of the secure KEM-based key encapsulation that the suggested system is based on.

iii. Forward secrecy and key compromise When employed that way, Kyber KEM encapsulation is ephemeral per session. The combination of a fresh encapsulation per session and session key rotation achieves forward secrecy in messages that are encrypted after the key rotation.

Recommendation: Replace the simpler hashing step with a standard HKDF (HMAC-based extract-and-expand) construction. HKDF allows for the incorporation of salt and context values and it is the recommended key-derivation mechanism in modern protocols.

Figure 11: Example using liboqs

Figure 11 shows successful execution of the integrated Kyber + AES demo using liboqs confirms that the post-quantum keypair was generated, the shared secret was encapsulated and validated correctly, and AES encryption/decryption using the Kyber-derived key worked as expected. The output shows the ciphertext and the correctly recovered plaintext *"Hello,kyber world!"*, demonstrating that the hybrid PQC–symmetric workflow functions end-to-end.

Figure 12: key generation using Liboqs+NIST

Figure 12 demonstrates successful keypair generation, encapsulation, and decapsulation. The confirmation *"Success: Shared secrets match!"* verifies that both parties derived the same post quantum secure shared secret, proving that the Kyber key exchange is functioning correctly in the implemented environment.

## 5. Conclusion and Future Scope

This work aptly represents the design and implementation of an end-to-end encryption protocol based on quantum cryptography which conveys confidentiality, integrity, and privacy for digital communication. This protocol includes the use of Kyber for quantum key exchange, SHA-256 to derive a secure key, and XChaCha20 for symmetric encryption giving strong and efficient secure messaging between parties[10]. Following are the recommendations from the work.

- Implement Kyber (appropriate parameter set) for the session key establishment and derive AES keys using HKDF(SHA-256 with context).
- Implement AES-GCM for authenticated encryption and be careful to manage nonce accordingly.
- Leverage amortization of the costs of using Kyber across sessions to keep latency to a minimum on small messages.
- Transition to optimized/native implementations and utilize hardware crypto (AES-NI) in production.
- Include robust logging, offline key-protecting measure, and rekeying policies to maintain forward secrecy and key hygiene. In addition, perform benchmarks for networked and concurrent production readiness testing.

This work concludes that a Hybrid Quantum-Classical End-to-End Encryption System is doable. Combining Kyber (for post-quantum key encapsulation), AES (symmetric) encryption, and SHA-256 hashing (for key making and integrity checks) makes things private, confirms the sender, and keeps communications secure going forward. The system also stands up against quantum computer attacks on older encryption.

This setup focuses on clear communication, speed, and encrypting in real-time. It works well for secure messaging and IoT setups. The design can be changed to fit new crypto updates and run on different platforms using containers. Future work can focus on group messaging, IoT improvement, etc.

**Source code:**

https://github.com/AditiGandhi2004/Integrate_Encryption